\documentclass[12pt]{iopart}

\usepackage{graphicx,color,soul}

\usepackage{iopams}

\newcommand{\bra}[1]{\left\langle{#1}\right|}

\begin{document}

\title{Dangling-bond charge qubit on a silicon surface}

\author{Lucian Livadaru$^{1,2}$, Peng Xue$^{3,4}$, Zahra Shaterzadeh-Yazdi$^3$, Gino A DiLabio$^1$, Josh Mutus$^2$, Jason L Pitters$^1$, Barry C Sanders$^3$ and Robert A Wolkow$^{1,2}$}
\address{$^1$ National Institute for Nanotechnology, National Research Council of Canada, Edmonton, Alberta T6G 2M9, Canada}
\address{$^2$ Department of Physics, University of Alberta, Edmonton, Alberta T6G 2J1, Canada}
\address{$^3$ Institute for Quantum Information Science, University of Calgary, Alberta T2N 1N4, Canada}
\address{$^4$ Department of Physics, Southeast University, Nanjing 211189, P. R. China}
\ead{lucian@ualberta.ca}

\begin{abstract}
Two closely spaced dangling bonds positioned on a silicon surface and sharing an excess electron are revealed to be a strong candidate for a charge qubit.
Based on our study of the coherent dynamics of this qubit, its extremely high tunneling rate $\sim \!\! 10^{14}{\rm s}^{-1}$ greatly exceeds the expected decoherence rates for a silicon-based system, thereby overcoming a critical obstacle of charge qubit quantum computing.
We investigate possible configurations of dangling bond qubits for quantum computing devices.
A first-order analysis of coherent dynamics of dangling bonds shows promise in this respect. 

\end{abstract}


\maketitle
\date{today}

\section{Introduction}
Quantum computing (QC) enables certain problems to be solved much faster than by known classical algorithms~\cite{Gro97}, and certain quantum algorithms are believed to speed up exponentially solving other problems such as factorization~\cite{Sho94}. Semiconductor solid-state implementations, especially in silicon, are particularly attractive because of the advanced state of silicon technology and the desire to integrate standard silicon-chip computing with quantum computation. Silicon-based qubits could be manifested as nuclear spin~\cite{Kan98}, electron spin~\cite{LD97,vrijen2000electron,barrett1,sanders2008visualizing}, and charge qubits~\cite{HFC+03,GHW05}. Although charge qubits have been successfully created in superconducting Cooper pair boxes~\cite{NPT99,Wallraff1}, realizations of semiconductor charge qubits are difficult because of strong decoherence effects.
Electron-spin qubits offer an alternative approach but face severe challenges such as readout:
in fact a promising approach to reading spin qubits first converts them to charge qubits~\cite{EHW+04}.
Thus, semiconductor charge qubits are important either as quantum information carriers or as intermediaries for spin-qubit readout. 

The charge qubit is manifested as a quantum dot pair such that having an excess electron on the `left' (or `right') dot corresponds to the logical state $\left|0\right\rangle$ (or the orthogonal state $\left|1\right\rangle$).
Coherent tunneling between the two quantum dots with a tunneling frequency $\Delta$ yields superpositions of $\left|0\right\rangle$ and~$\left|1\right\rangle$.
These states correspond to position encoding, whereas the symmetric and antisymmetric qubit states correspond to energy encoding.

Here we show that a pair of quantum dots, with a dot corresponding to a silicon-surface dangling bond~(DB), should be an excellent semiconductor charge qubit with low decoherence. Such pairs of coupled DBs have recently been fabricated~\cite{HBD+09}. As previously proposed semiconductor charge qubits critically suffer from large decoherence~\cite{FHH04,barrett1}, it appears worth investigating whether this obstacle can be tackled by shrinking the scale of the constituent quantum dots, concomitantly  with the spacing between them.
The motivation for shrinking the size is that the tunneling rate increases exponentially with decreasing inter-dot separation whereas, as shown below, decoherence is expected to scale weakly with inter-dot separation.
Therefore, decreasing separation allows many coherent oscillations before the onset of significant decoherence. 

Implementing such a strategy involves abandoning heterostructure quantum dots and instead adopting DB quantum dots on an H-terminated Si surface.
As this quantum dot size is of atomic dimensions, commensurately close spacing of dots is enabled~\cite{HBD+09}. Pairs of appropriately separated DBs at low temperature share precisely one excess electron per pair (denoted DB-DB$^-$), yielding coherent tunnel coupling between the two DBs~\cite{HBD+09}.
This fact strongly motivates opening a new area of interest, namely coherent dynamics of tunnel-coupled DBs on a silicon surface with the potential of exploiting them in quantum information processing. 

The novelty of our qubit consists is its extremely high tunneling rate $\Delta \sim \!\! 4.7\times10^{14}{\rm s}^{-1}$, compared to a maximum order-of-magnitude of $10^{12}{\rm s}^{-1}$ in previously proposed charge qubits of atomic scale~\cite{barrett1}.
Another novel feature is the relative ease of fabrication, as already demonstrated~\cite{HBD+09}. DB quantum dots can be separated by subnanometer distances, are almost physically identical, and, as surface entities, are directly amenable to measurement and control. Furthermore, unlike the case for quantum dots composed of atoms buried in bulk media, the silicon-based scheme proposed here trades the extraordinarily difficult requirement of precisely positioned single dopant atoms with the challenging but attainable requirement that single H atoms be removed by a scanned probe. These DB-DB$^-$ charge qubits could form the basic units of a quantum computer, the dynamics of which we describe with the extended Hubbard model~\cite{Hub78}. 

The outline of our paper is as follows. In Section~\ref{sect1}, we describe the physical characteristics of the silicon dangling bonds and we show that all DB-DB$^-$ pairs can be initialized such that each excess electron is either in the `left' or the `right' DB of each pair; subsequently the potential landscape can be tilted so that all are initialized in the `left' state. In Section~\ref{sect2} we formulate the quantum dynamics of a system of DBs in the frame of an extended Hubbard model. In Section~\ref{sect3} we analyze the decoherence effects on the quantum dynamics of the DB system due to its interaction with external factors such as thermal noise in conductors and phonons in the silicon substrate. Without undertaking a full analysis, we mention in Section~\ref{sect4} how these DB-DB$^-$ pairs could find applicability in a quantum computer circuit e.g.\ the flying-qubit circuit model based on a bulk-silicon electron-spin qubit version~\cite{HGFW06} or on measurement-based one-way quantum computing~\cite{RB01}. The complete fulfillment of the DiVincenzo criteria is not being explicitly addressed, as it exceeds the scope of this paper. 

\section{Dangling bond pairs as charge qubits}\label{sect1}
\begin{figure}
\centering
\begin{tabular}{cc}
 \includegraphics[width=0.55\linewidth,clip=]{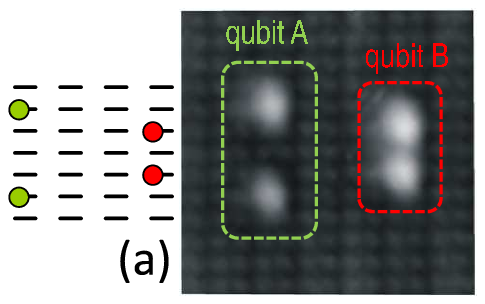} & 
 \includegraphics[width=0.35\linewidth,clip=]{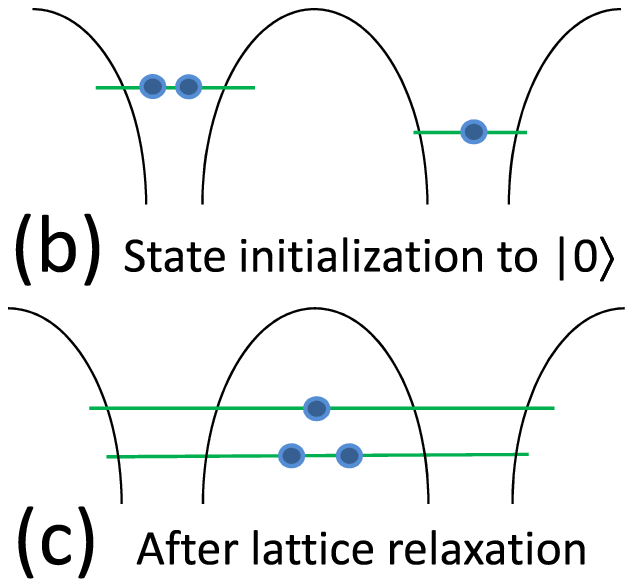}
\end{tabular}
 \caption{
	(a) Variably spaced qubits in an atom-resolved STM image (46\AA$\times$46\AA, 2V, 0.2nA) created from pairs of DBs on a H-Si(100)2$\times$1 surface, separated 
	by 15.36\AA~(qubit A) and 7.68\AA~(qubit B). Dangling bonds appear as bright protrusions in the gray scale image. A schematic (left) shows the 
	position of  DBs (red and green circles) on the Si surface.  
	Black dashes represent silicon dimers.
 (b) A DB-DB$^-$ pair modeled as double-well potential, with the extra electron 
 at the left well immediately after initialization to $\left|0\right\rangle$.
 (c) Relaxed ground state of the DB electrons after lattice 
 relaxation has completed.
	}
 \label{fig:qubit}
\end{figure}

A neutral DB hosts a bound electron within the Si $1.1$eV bulk band gap. The itinerant electrons available in a doped semiconductor can provide a second electron of opposite spin to the DB, thus rendering it a DB$^-$. If two DBs are sufficiently close together ($\leq16$\AA), Coulomb repulsion ensures that a doubly-charged DB$^-$-DB$^-$ pair cannot form~\cite{HBD+09}. Hence, a closely-spaced DB pair shares one extra electron tunneling between two centers, suggesting its use as a charge qubit.
Tunnel-coupled DBs, as shown in Fig.~\ref{fig:qubit}(a), have been created on a Si(100) surface by first passivating a Si(100) surface with a hydrogen monolayer then using a scanning tunneling microscope (STM) tip to remove H atoms at selected sites~\cite{HBD+09}. The separation between the two DBs forming a pair has a strict lower bound of $3.84$\AA\ as determined by the lattice spacing of the Si(100) surface, whereas the upper bound for enabling a qubit is given by the tunneling range of about $16$\AA. Distinct pairs are created farther apart than this limit to avoid inter-pair tunnel coupling. Here we claim that DB-DB$^-$ pairs exhibit coherent quantum dynamics and can serve as good charge qubits.


%
The localized nature of the DB wavefunction and its energy level in the band gap allows us to formulate an electron-confinement model corresponding to a 
potential well accounting for the effect of the environment. 
Such a potential well description must render the correct eigenstate energy 
and orbital size, and must allow for electron excitation into the bulk conduction band of the crystal.
For a neutral DB, we calculate the binding energy of an electron to be about 0.77eV~\cite{gaussian}.
In a highly-doped n-type crystal, a high Fermi level  of the crystal allows an extra electron to be localized at a DB, rendering the DB site negatively charged;
similarly, if the crystal is p-type, the DB can lose all its electrons thereby becoming positively charged.

This localization has two important physical consequences: 
\begin{itemize}
	\item a 0.5eV upward shift of the DB$^-$ energy level relative to that of a neutral DB's to 
		$\sim$0.85eV above the valence band edge
		(a change in the potential well resulting in weaker confinement and a lower ionization energy);
 \item a local lattice deformation whereby the host Si atom at a DB$^-$ 
		is raised by 0.3\AA~ from the plane of the surface. 
		After the electron tunnels out of a DB$^-$, the lattice begins to relax.
\end{itemize}

In Figs.~\ref{fig:qubit}(b,c) we depict a DB pair as an effective 
double-well potential with~(b) an excess electron at the left well 
immediately after release from a biasing external field, as required 
for qubit initialization, and~(c) after complete lattice equilibration 
when the potential landscape becomes symmetrical. 
Due to the localized extra charge, 
the double-well in case~(b) does not exhibit the symmetry of case~(c), 
and the DB energy is shifted upward at the left site.
Consequently, during lattice relaxation, the coherent oscillation 
between the two DBs takes place between 
two wells of slightly different shapes, resulting in a periodic oscillation 
that is biased towards the `left' (excess electron spends more time
on the left than on the `right'). 
Slow relaxation of the lattice will modify the electron oscillation and cause weak decoherence commensurate with the ratio of relaxation rate to oscillation rate.

We calculate tunneling rates in a DB-DB$^-$ pair for various separations by two different methods.
For DB separations of $3.84$\AA\ and~$7.68$\AA, tunnel splitting is determined to be $307.7$meV and~$87.8$meV, respectively, by time-dependent density-functional theory on cluster models~\cite{gaussian}.
These correspond to tunneling rates of 4.67$\times 10^{14}{\rm s}^{-1}$  and 1.33$\times 10^{14}{\rm s}^{-1}$, respectively.
For greater separations the size of the silicon cluster model becomes prohibitively expensive for this computation,
and we resort to simpler approximations, namely the Wentzel-Kramers-Brillouin (WKB) method.
The results are plotted in Fig.~\ref{rates1}.

\section{Quantum dynamics of DB system}\label{sect2}

\subsection{Hamiltonian dynamics}
Our estimated decoherence rates (Section~\ref{sect3}) are orders of magnitude smaller than tunneling rates, for chosen intra-qubit DB separations.
Therefore, the dynamics of DBs on the surface can be described by a Hamiltonian~$\hat{H}$
that acts upon the Hilbert space spanned by zero, one, or two electrons
at each DB upon the silicon surface.
On-site energy, electron tunneling (hopping), intra- and inter-DB Coulomb repulsion between electrons, and potential differences across the surface are all incorporated into~$\hat{H}$.

We consider any number of DBs on the surface, with~$i$ labeling the DB site.
Let~$E_{\rm{os}}$ be the on-site energy of an electron at any DB,
which includes a constant surface chemical potential offset,
and~$\eta_i$ be a site-dependent energy correction due to local field effects.
The slow lattice deformation due to the excess electron
and the potential well deformation due to external biasing fields
can be incorporated into this $\eta_i$ parameter.
The hopping integral between sites~$i$ and~$j$ is~$T_{ij}= \hbar \Delta_{ij}/2$,
which depends on the separation~$r_{ij}$ between the two DBs.
$U_i$ denotes the energy  cost of putting two electrons of opposite spin at the same site~$i$, including the screening energy. The cost of putting one electron with spin~$\sigma\in\{\uparrow,\downarrow\}$ at site~$i$ and another electron of spin~$\sigma'$ at site~$j$ is denoted~$W_{i\sigma j\sigma'}$.

Tunneling between DB sites can be controlled by modifying the inter-site potential bias. For example two sites~$i$ and~$j$ can have a time-dependent potential difference of~$V_{ij}(t)$.
For $\hat{c}_{i,\sigma}$ ($\hat{c}^\dagger_{i,\sigma}$) the annihilation (creation) operator
for an electron with spin~$\sigma$ at site~$i$ and
$\hat{n}_{i,\sigma}=\hat{c}^\dagger_{i,\sigma}\hat{c}_{i,\sigma}$ the number operator
for electrons of spin~$\sigma$ at site~$i$,
the potential difference operator between sites~$i$ and~$j$ is
\begin{equation}
	\hat{V}\equiv\frac{1}{2}\sum_{i<j,\sigma}V_{ij}(\hat{n}_{i,\sigma}-\hat{n}_{j,\sigma}).
\end{equation}

We now have all the terms required to express the Hamiltonian as
an extended Hubbard model~\cite{Hub78}:
\begin{eqnarray}
\label{eq:generalHam}
	\hat{H}
		&=& \sum_{i,\sigma}(E_{\rm{os}}+\eta_i)\hat{n}_{i,\sigma}	
		-\sum_{\stackrel{i<j}{\sigma}}T_{ij}(\hat{c}^\dagger_{i,\sigma}\hat{c}_{j,\sigma}  
		+\hat{c}^\dagger_{j,\sigma}\hat{c}_{i,\sigma})	\nonumber	\\	&&
		+\sum_i U_i\hat{n}_{i,\uparrow}\hat{n}_{i,\downarrow}
		+\sum_{i<j,\sigma,\sigma'}W_{i\sigma j\sigma'}\hat{n}_{i,\sigma}\hat{n}_{j,\sigma'}+\hat{V}.
\end{eqnarray}
As DB-DB$^-$ qubit tunneling is much faster than decoherence processes,
Hamiltonian~(\ref{eq:generalHam}) is justified by working in a regime of coupled qubits with standard descriptions of Markovian qubit decoherence~\cite{CL81}.

Typical values of system parameters are:
		the Fermi level (or more exactly, the chemical potential) $E_F= 0.95$eV, for a medium-level n-type doped silicon sample; 
		the neutral DB energy level $E_{DB}= 0.35$eV, 
		the negative DB energy level $E_{DB-}= 0.85$eV 
		(energy values given with respect to the silicon valence band edge). 
We extracted the Hubbard model parameters from the results of our \emph{ab initio} calculations:
$E_{\rm{os}}= 0.52$eV and~$U=1.00$eV. For a DB separation of 3.84\AA, $T=0.154$eV, and $W=0.37$eV.

\subsection{Qubit dynamics}
Hamiltonian~(\ref{eq:generalHam}) describes dynamics for quite general configurations of DBs on the silicon surface. For quantum computing, we need to generate entanglement by applying time-dependent gate potentials for specific qubit separation and relative orientation on the Si surface.
A highly ordered pattern of DBs, corresponding to grouping DBs into pairs of nearby DBs, and well chosen separations between pairs, greatly simplifies~(\ref{eq:generalHam}). For DBs on the silicon surface, electron spin is preserved so can be neglected; hence the `left' state~$\left|0\right\rangle$ and `right' state~$\left|1\right\rangle$ form a qubit basis with conjugate (energy) basis corresponding to
$\left|\pm\right\rangle=\left(\left|0\right\rangle\pm\left|1\right\rangle\right)/\sqrt{2}$. 
In fact the wave functions corresponding to `left' and `right' occupation are not	completely orthogonal, but the overlap is negligibly small for the case of two DBs separated by several \AA.

For $N$ DB-DB$^-$ pairs, the Hamiltonian can be conveniently rewritten in the qubit basis as a linear combination of quantum gates and tensor products thereof:
$\mathbf{1}$, $\hat{X}=\left|0\right\rangle\bra{1}+\left|1\right\rangle\bra{0}$
and~$\hat{Z}=\left|0\right\rangle\bra{0}-\left|1\right\rangle\bra{1}$.
Whereas~$i,j$ designates DB sites,
$\imath,\jmath$ denotes DB \emph{pair} sites
(equivalently charge qubits).
The Hamiltonian is now expressed as an operator sum that acts on and
between DB-DB$^-$ pairs:

\begin{equation}
\hat{H}_{\rm{q}}(t)
		=\kappa\mathbf{1}+ \sum_{\imath=1}^N\left[T\hat{X}_\imath
			+\frac{1}{2}\Delta V_\imath(t)\hat{Z}_\imath	+\frac{1}{2}\sum_{\jmath<\imath}W_{\imath\jmath}^{-}\hat{Z}_\imath\otimes\hat{Z}_\jmath\right].
\end{equation}
Intra-qubit separation is constant for all qubits
with~$U_0$ and~$W_0$ on-site and inter-site Coulomb interaction within each DB-DB$^-$ qubit,
and~$W_{\imath\jmath}$ is the inter-qubit Coulomb repulsion
$W_{\imath\jmath}^{\pm}=W_{\imath\jmath}^{\rm{s}}\pm W_{\imath\jmath}^{\rm{c}}$
and~$W_{\imath\jmath}^{\rm{s}}$ ($W_{\imath\jmath}^{\rm{c}}$) is
the inter-site Coulomb interaction between the same (cross) sites
of two DB-DB$^-$ pairs $\imath$ and~$\jmath$.
Then
\begin{equation}
	\kappa=N(3E_{\rm{os}}+3\eta+U_0+2W_0)+\frac{9}{2}\sum_{\imath<\jmath}^NW^+_{\imath\jmath}
\end{equation}
Qubit-specific time-dependent potential-landscape tilting~$\Delta V_\imath(t)$
is incorporated into $\hat{H}_{\rm{q}}(t)$, with~$T$ the intra-qubit tunnel splitting energy. 

\section{Decoherence analysis for DB-DB$^-$ qubit systems}\label{sect3}

For a DB-DB$^-$ qubit on H-Si(001) surface, we treat the decoherence mechanism due to various interactions with the environment within the spin-boson model.
The spin-boson model is a well-established simplified model for a few-level quantum system interacting with a bath of harmonic oscillators and for a two-level system, 
spin-boson dynamics has been extensively studied~\cite{leggett3}.
The spin-boson model has been previously employed for the treatment of decoherence in the P-P$^{+}$ charge qubit in bulk silicon~\cite{barrett1}. Based on earlier studies on silicon systems\cite{barrett1,andresen1}, we estimate that the main sources of decoherence for our system are: (i)~the voltage-fluctuations on the gate electrodes, and (ii)~the interaction between the qubit electron and phonons in silicon bulk and at the surface.
We discuss these sources below and calculate the corresponding decoherence rates for a DB-DB$^-$ qubit. 
Other decoherence sources, such as control errors are not included in this analysis, as they depend on a specific architecture of the device.
Decoherence due to stray charges in the system is also believed to be small\cite{HBD+09}, as the spacing between DBs in a qubit is much smaller than distances to the nearest trapped charges in the semiconductor.

The spin-boson Hamiltonian~\cite{leggett3} for a single qubit interacting with its environment is given by 
\begin{equation}
	\hat{H}_{\rm sb}=\hat{H}_{\rm qb}+\hat{H}_{\rm bath}+\hat{H}_{\rm int}
\end{equation}%
for~$\hat{H}_{\rm qb}$ and~$\hat{H}_{\rm bath}$ separate qubit and bath Hamiltonians, respectively, and
$\hat{H}_{\rm int}$ the interaction term. The latter is given by
\begin{equation} 
	\hat{H}_{\rm int}
		=\frac{1}{2}\hbar \hat{Z}\sum_i\lambda_i(\hat{a}_i^{\dagger}+\hat{a}_i)
		=\frac{1}{2}\hat{Z} d \sum_ic_i\hat{x}_i.
\label{inter2}
\end{equation}
where $i$ denotes a harmonic oscillator mode with frequency $\omega_i$,
and~$\hat{a}_i^{\dagger}$ and~$\hat{a}_i$ are the creation and anihilation operators for mode~$i$ within the second quantization formalism, and~$\lambda_i$ is the coupling coefficient between the qubit and mode~$i$. 
In this model, the coupling between the qubit and the bath depends linearly on the coordinate of the qubit and those of the harmonic oscillator modes. This is obvious in the expression in Eq.~(\ref{inter2}),
where $d$ is the distance between the localized qubit states, $\hat{x}_i$ is
the spatial coordinate of mode~$i$, and~$c_i$ is the coupling strength
between the qubit and mode~$i$.

Earlier studies show that, for any system characterized by the equilibrium statistical average over the initial and
final states of the bath, the only physically relevant quantity in the spin-boson model
is the so-called spectral density function of the bath~\cite{leggett3}
\begin{equation}
	J(\omega )=\frac{\pi }{2}\sum\limits_i\delta (\omega -\omega_i)\frac{c_i^2}{m_i\omega_i}.
\label{spectral1}
\end{equation}
A large class of open systems can be characterized by a spectral function of the form%
\begin{equation}
J(\omega )=\alpha \omega ^{s}\exp (-\omega /\omega_{\rm c})
\label{spectral2}
\end{equation}%
for~$\omega_{\rm c}$ a cutoff frequency and~$\alpha$ and~$s$ empirically-fitted constants. 
For $s=1$, the bath is said to be ohmic.
In our study, we assume the spectral density to be of the form~(\ref{spectral2}), for which we will specify appropriate parameters $\alpha$, $s$, and~$\omega_{\rm c}$.

The spin-boson model, although one of the simplest dissipative two-state systems,
does not have a general analytic solution.
The dynamical behavior of this model essentially depends on the ratios between the parameters $\Delta$, 
$\omega_{\rm c}$, and~$k\Theta$.
For practical purposes, the most common solutions are perturbative ones (in which the weakest term in the total Hamiltonian plays the role of the perturbation) and path integral techniques.
For example, in the adiabatic limit, $\Delta \gg\omega_{\rm c}$, the bath evolves quite slowly and has an almost classical behavior, whereas in the non-adiabatic limit, $\Delta\approx \omega_{\rm c}$,
the golden rule offers a reliable solution.
Other limiting cases also involving~$k\Theta$ and other energy scales are well understood~\cite{garg6,dak2}.

As the bare tunneling rate in a qubit increases, fluctuations in the tunneling splitting can play an important role in the coupling with the environment. This can be described as terms in the interaction Hamiltonian proportional to $\sigma_x$ and~$\sigma_y$, which do not explicitly appear in the spin-boson Hamiltonian.
However, it was shown in early studies on the spin-boson model~\cite{leggett3} that this effect can be still accommodated by the spin-boson model via renormalizing the bare tunneling rate and the applied bias.
The only condition is that the tunneling rate be much less than the classical oscillation frequency $\omega_0$ corresponding to electron confinement in an isolated DB.
As the confinement energy in a DB is about 0.6eV, this condition is generally fulfilled for all qubit configurations, with the exception of the qubit with a separation of 3.84\AA.
Therefore, we must bear in mind that, in this limit, the accuracy of the spin-boson model may be unreliable.

\subsection{Decoherence due to Johnson-Nyquist voltage fluctuations}

Johnson-Nyquist noise (also known as Johnson noise) is due to random thermal
fluctuations of the charge carriers in a conductor or semiconductor. For the
purpose of calculating its effect on the coherent oscillations in a charge
qubit, we employ the spin-boson model with the qubit being the two-level
system and the gate electrode being the bath. In order to obtain a reliable
estimate of the decoherence effect, we need to look at how the power
spectrum of the bath compares to the bare tunneling frequency of the
qubit.

The Johnson noise stretches uniformly in the frequency range from zero to
the quantum limit of $k\Theta/\hbar $ (where $k$ is the Boltzmann constant and $\Theta$ is the temperature), which practically means up to about $10^{11}$-$10^{13}{\rm s}^{-1}$. 
In particular, we can see that for a temperature $\Theta=$ 4 K the spectrum  has a cutoff frequency $\omega_{\rm c}=5.2\times 10^{11}{\rm s}^{-1}$. As the bare tunneling frequency of the charge qubit in our dangling bond implementation has a value $\Delta \approx 10^{14}{\rm s}^{-1}$ $ \gg \omega_{\rm c} \approx 10^{12}{\rm s}^{-1}$, we can safely regard the effect of the bath on the qubit as being approximately adiabatic (potential energy changes experienced by a qubit electron due to fluctuations in the bath vary slowly in time compared to the bare tunneling frequency of the qubit).

To proceed we assume ohmic dissipation corresponding to~$s=1$ in Eq.~(\ref{spectral2})
so the spectral function in the spin-boson model has the form 
\begin{equation}
G(\omega )=\alpha_{\rm JN} \omega \exp (-\omega /\omega_{\rm c})
\label{G1}
\end{equation}%
for
\begin{equation}
	\alpha_{\rm JN} =\frac{\eta d}{2\pi\hbar}.
\end{equation}
a dimensionless dissipation/coupling function of the
distance between the two charge centers~$d$ and viscosity coefficient $\eta$.

One of the simplifying features of the spin-boson Hamiltonian is that, in the limit of weak qubit-bath coupling, the decoherence times $T_1$ (describing population decay) and~$T_2$ (describing coherence decay) are equal to second order in the coupling~\cite{leggett3}, which allows to characterize the system by a single decoherence rate $\Gamma= 1/T_1$.   
In the adiabatic limit ($\Delta\gg\omega_{\rm c}$), and for the case when 
$\omega_{\rm c} \gg \Gamma$ (which can be verified a posteriori), the decoherence rate for the qubit can be determined according to~\cite{dak2,garg6} 
\begin{equation}
	\Gamma_{\rm JN}
		=\frac{1}{2}\frac{\sqrt{\pi }\hbar\Delta^2}
			{1+\frac{\hbar\Delta^2}{\omega_{\rm c}E_{\rm r}}}
			\frac{\exp (-\frac{E_{\rm r}}{4kT})}{\sqrt{\frac{E_{\rm r}}{k\Theta}}}
\label{Gamma1}
\end{equation}
for~$E_{\rm r}$ the bath reorganization energy
\begin{equation}
	E_{\rm{r}}=\hbar \int\limits_0^\infty {\rm d}\omega \frac{G(\omega )}{\omega }
\end{equation}
which can be calculated from Eq.~(\ref{G1}) to yield
\begin{equation}
	E_{\rm r}=2\alpha_{\rm JN} \hbar \omega_{\rm c}.
\end{equation}

\begin{figure}[tb]
\begin{center}
	\includegraphics[width=0.7\linewidth,clip=]{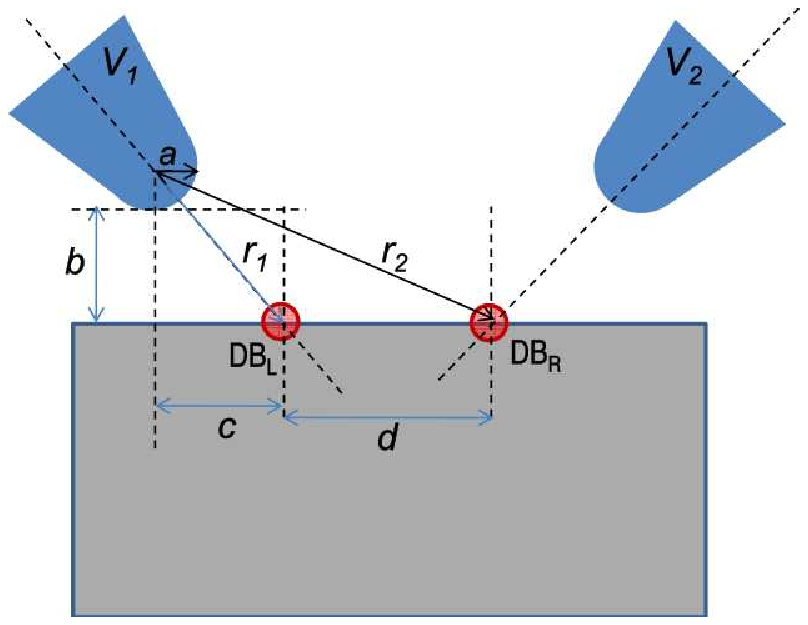}
\end{center}
	\caption{Sketch of the gating geometry for our proposed DB-DB- qubit on silicon surface. DBs are indicated as red circles and are indexed L and R corresponding to their locations.
	The electrodes (based on STM tips) are indicated in blue and have fixed potentials $V_1$ and~$V_2$, with $V_{12}= V_1- V_2$. The radius of the electrode at the apex is $a$.}
	\label{qubit_gating}
\end{figure}

For a typical charge-qubit gating~\cite{barrett1}
\begin{equation}
	\alpha_{\rm JN}= \frac {e^2\beta^2 R_{\rm g}}{4h}
\label{alp1}
\end{equation}
for $R_{\rm g}$ the resistance of the gate circuit,
and
\begin{equation}
	 \beta =\frac{\delta V_{\rm LR} }{\delta V_{\rm 12}}   
\end{equation}
is another dimensionless parameter that depends solely on the system geometry.
Here~$V_{\rm LR}$ the difference between the electrostatic potentials at the L and R sites
and~$V_{\rm 12}$ the difference between in the applied voltage on the two electrodes.

A simple approximation (but yielding good order-of-magnitude estimate) for the electrostatic problem
(Fig.~\ref{qubit_gating}) yields
\begin{equation}
	 \delta V_{\rm LR} =\left(\frac{a}{r_{1} } -\frac{a}{r_{2} } \right)(V_{1} -V_{2}) 
\end{equation}
whence we obtain 
\begin{equation}
	\beta_{\rm JN} =a\left(\frac{1}{\sqrt{c^2 +(a+b)^2}} -\frac{1}{\sqrt{(c+d)^2 +(a+b)^2}} \right).
\end{equation}
Plugging in reasonable estimates for the parameters: $a=b=c=2$nm, $d=0.772$nm, we find $\beta=0.036$.
Further, by assuming~$R_g= 50\Omega$  and using Eq.~(\ref{alp1}) yields
$\alpha_{\rm JN}=6.364\times 10^{-7}$.

Finally we can calculate the decoherence rate due to Johnson-Nyquist noise for a DB-DB$^-$ 
charge-qubit implementation depicted in Fig.~\ref{qubit_gating} with typical parameters $\Delta=1.33\times 10^{14}{\rm s}^{-1}$, $\omega_{\rm c}=1.31\times10^{11}{\rm s}^{-1}$ ($\Theta$= 1 K).
From Eq.~(\ref{Gamma1}) we obtain
\begin{equation}
	\Gamma_{\rm JN}= 1.30\times 10^8{\rm s}^{-1},
\end{equation}
which is much less than $\omega_{\rm c}$, thereby showing that our approximations are consistent.
Note that, as discussed above, the spin-boson model is less reliable for high qubit-tunneling rates,
which means that its results are unreliable for the closest DB separation of $3.84$\AA.
Nonetheless, the decoherence for all other DB separations can be accurately treated by this model because the corresponding tunnel-splitting energy is much less than the binding energy~$0.6$eV.

We claim that our decoherence rate is much smaller than the bare tunneling frequency of the qubit, a very favorable fact for implementing reliable quantum gates.
This also compares well with the decoherence rate due to Johnson noise in the P-P$^+$ charge qubit implementation proposed in previous studies~\cite{barrett1,hollenberg2}. Notice however that the treatment of decoherence due to Johnson-Nyquist noise in our DB system is quite different from the P-P$^+$ system.
This is due to the fact that in the latter system, the bare tunneling frequency $\Delta$ is actually much smaller than the cutoff frequency of the bath $\omega_{\rm c}$, which requires different approximations to be employed when calculating decoherence rates.

\begin{figure}[tbp]
 \includegraphics[width=0.8\linewidth,clip=]{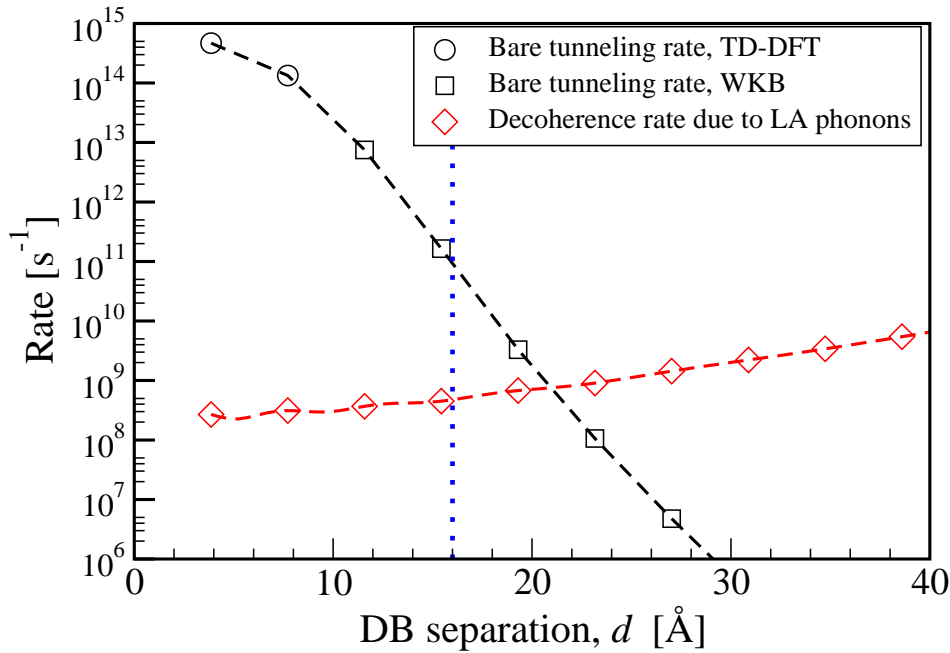}
 \caption{
	Bare tunneling rates of the excess electron in a charge qubit 
	by time-dependent density-functional theory (black circles) 
	and the WKB method (black squares) vs. DB separation $d$.
	The red line depicts the calculated 
	decoherence rate due to longitudinal-acoustical (LA) phonons. 
	The vertical blue dotted line indicates (to its left) the region in which DBs are tunnel coupled.
	}
 \label{rates1}
\end{figure}
\subsection{Decoherence due to electron-phonon interaction}

Previous studies on electron-phonon scattering in reduced-dimension systems have found that,
for zero-dimensional systems, the scattering rates are smaller by at least an order-of-magnitude than in one- and two-dimensional systems~\cite{bockelmann1}. This is due to the fact that, for a given initial state of the electron, the number of final states is greatly reduced in the zero-dimensional case.
For our system, if the DB-DB$^-$ charge qubit is in the anti-symmetric state~$|-\rangle$,
only the (symmetric) ground state~$|+\rangle$ is lower in energy, thereby drastically reducing coupling to phonons.

Nonetheless, for our system, the interaction between electrons and phonons can be a serious source of decoherence, and we anticipate that in our system it dominates all other forms.
From previous experimental and theoretical studies~\cite{tutuncu1} on phonons in the Si(001) crystal, we know that the phonon spectrum can extend up to about 70 meV, corresponding to a frequency of $1.06\times 10^{14}{\rm s}^{-1}$. This rate is comparable to the bare tunneling frequency for the charge qubit, which means that the adiabatic approximation used in the previous section fails.
A different approach is required and, as in previous theoretical analyses of the electron-phonon interaction, we calculate the rates of electron-phonon scattering within the frame of the first-order perturbation theory via the Fermi golden rule~\cite{bockelmann1}
\begin{eqnarray}
	\Gamma _{\rm e-ph}
		&=& \frac{2\pi }{\hbar } \sum _{f,\bi{q}}\alpha^2(\bi{q})
			\left|\left\langle \psi_f |e^{\pm i \bi{\bi q}\cdot \bi{r}}|\psi_i\right\rangle\right|^2
				\nonumber	\\	&&
			\times\delta (E_f -E_i \mp E_\bi{q} ) 
\left[ n_{B} (E_{q} ,\Theta)+\frac{1}{2} \mp \frac{1}{2} \right]  
\end{eqnarray}
for~$i$ and~$f$ indices for the initial and final electronic states,
$\bi{q}$ the phonon wavevector, $E_q$ the phonon energy,
$\alpha(q)$ a coupling function,
$n_B$ is the Bose occupation distribution, 
and upper/lower signs corresponds to absorption/emission of a phonon by the qubit.

Below, we quantify the coupling of the charge qubit with the acoustic phonons only. The coupling of electrons to the \textit{longitudinal-optical} (LO) phonons is also possible.
However, optical phonons have a more discrete-like energy spectrum (set of distinct spectral lines), and this fact prevents any first-order coupling to electrons, unless the energy matching condition
\begin{equation}
	\hbar \omega_{\rm LO} =E_f -E_i
\label{eq:LO}
\end{equation}
is fulfilled.
Condition~(\ref{eq:LO}) can be avoided in our system by judiciously choosing the inter-dot distance and the amplitude of the applied bias. 

If the coupling is given via a deformation potential, $D$, then the coupling function above can be shown to be
\begin{equation}
	 \alpha ^2 (q)=\frac{D^2 }{2\rho c_{s}^2 \Omega } \hbar c_{s}^2 q
\end{equation}
where $c_{s}^2$ is the longitudinal sound velocity, $\rho$ is the density, and~$\Omega$ is a normalization volume. Piezoelectric coupling to acoustic phonons is also possible, but in general it is much weaker (by an order of magnitude~\cite{bockelmann1}) than the coupling via a deformation potential.
After appropriate manipulation, the expression for the scattering rate can be reduced to~\cite{benisty1}
\begin{equation}
	\Gamma_{\rm e-ph} =\frac{D^2 q_{if}^{3} }{8\pi ^2 \hbar \rho c_{s}^2 } \left[n_{B} (E_{q} ,\Theta)+\frac{1}{2} \mp \frac{1}{2} \right]\int {\rm d}\Omega _\bi{q} |\left\langle \psi_f |e^{\pm i \bi{q}\cdot \bi{r}} |\psi_i \right\rangle |^2    
\end{equation}
where~$q_{if}=E_{if} /\hbar c_s$ for~$E_{if}$ the energy difference between the~$i$ and~$f$ states
and~${\rm d}\Omega _\bi{q} $ is the solid angle element in $\bi{q}$-space.

As in previous studies~\cite{barrett1,andresen1} we assume that a DB can be modeled as a 1s hydrogen-like orbital with a renormalized Bohr radius, $a_B$,
and we fit this parameter so that the tunnel splitting of a DB pair derived from the hydrogen-like model reproduces the value predicted by our \emph{ab initio} calculations for a DB separation of $7.68$\AA.
Then it can be shown that the rate of phonon emission is given by 
\begin{equation}
	 \Gamma_{\rm e-ph}
	 	=\frac{64D^2q^3\sin^2\theta}{\pi\rho\hbar c_s}\frac{n_{B} (E,\Theta)+1}{ [(qa_{B} )^2 +4]^4}
			\left(1-\frac{\sin qd}{qd}\right) 
\end{equation} 
where~$\theta =\tan^{-1}(\hbar \Delta /\varepsilon)$, and~$\varepsilon $ is the applied bias on the qubit, and~$d$ is the dot separation.
Note that the results for~$\Gamma_{\rm e-ph}$ are of the same order of magnitude for any other form of the isolated dot wavefunction exhibiting exponential decay, as long as the decay rate is similar.

The decoherence rate as a function of dot separation is plotted in Fig.~\ref{rates1} together with the bare tunneling rates for different intra-qubit DB separation. For DB separations of $3.84$\AA\ and~$7.68$\AA, the tunneling rates ($4.67\times 10^{14}{\rm s}^{-1}$  and 1.33$\times 10^{14}{\rm s}^{-1}$, respectively) are denoted by circles,
and tunneling rates for greater DB separations are calculated by the Wentzel-Kramers-Brillouin (WKB) method.
A continuous line corresponding to interpolation between calculated tunneling rates for different 
inter-dot separations allows direct comparison

The black dashed line joining the calculated points shows an interpolation  of the results obtained by the two methods whereas, for comparison, the red curve in Fig.~\ref{rates1} shows the calculated decoherence rate due to electron interaction with LA phonons in the silicon crystal.

Note that, for our DB system, the above rate $\Gamma_{\rm e-ph} $ is greater than the decoherence rate due to Johnson noise in the electrodes, $\Gamma_{\rm JN}$, calculated in the previous section. Thus we identify $\Gamma_{\rm e-ph}$ as the dominant decoherence rate. We note an important fact for our qubit: relaxation via this mode occurs over several nanoseconds whereas the tunneling period for the DB-DB$^-$ pair with a few \AA\ separation is close to 10fs, which enables many coherent qubit oscillations before decoherence sets in. 

Other phonon modes both in bulk and at the surface~\cite{tutuncu1} are less likely to couple to electron tunneling due to their discrete-like energy spectrum. Without considering all the selection rules, at least for DB separations of $3.84$\AA\ and~$7.68$\AA, there are no phonon modes to match the tunnel splitting energy, as the highest phonon energy is about 70meV. A more detailed analysis of the qubit coupling to the optical phonon modes is beyond the scope of this paper. Overall, we estimate that for our closely spaced qubits Rabi-type oscillations will take place over many periods before the onset of critical decoherence, illustrating the advantage of closely spaced quantum dots.

\section{Applications to quantum computation}
\label{sect4}

For our DB-DB$^-$ pair to be an effective charge qubit, initialization to a simple pure state and qubit-specific readout are critical, as two of  DiVincenzo's five criteria~\cite{DiV00} (the other criteria are scalability, a universal set of gates, long coherence times, with coherence time addressed in the previous section).
Complete fulfillment of DiVincenzo's five criteria is beyond the scope of this work.
but here we briefly discuss coherence times, initialization, and readout. 

As shown in the sections above, the single-qubit gate time is of the order of $10^{-14}$s, whereas the two-qubit gate time can be estimated as  $h /(W_{\imath\jmath}^{\rm{s}}- W_{\imath\jmath}^{\rm{c}})$. 
Estimated values for the two-qubit gate times are given in Table~\ref{table1} for chosen DB configurations.
For example, for the first DB configuration in that table, the expected decoherence time is $2\times 10^{-8}$s,
thereby yielding an error probability of 2.3$\times 10^{-6}$ ($1.6\times 10^{-5}$) for the single qubit (two-qubit) gate, well within the tolerance demanded by standard quantum error correction protocols.

\begin{table}
\caption{\label{table1}Estimates of the single-qubit and two-qubit gate times for chosen DB configurations. $s_{\imath\jmath}^{\rm{inter}}$ is the distance between the two identical and parallel qubits, $d_{\rm{DB-DB}}^{\rm{intra}}$ is the intra-qubit DB-DB distance, and~$W_{\imath\jmath}^{\rm{s}}$ and~$W_{\imath\jmath}^{\rm{c}}$ are the Coulomb interactions as explained in the text.}
\begin{tabular}{@{}cccccc} 
\br
$s_{\imath\jmath}^{\rm{inter}}$ [\AA]  & $d_{\rm{DB-DB}}^{\rm{intra}}$ [\AA]  & 1-QB gate time [s]  & $W_{\imath\jmath}^{\rm{s}}$ [eV] & $W_{\imath\jmath}^{\rm{c}}$ [eV] & 2-QB gate time [s] \\

\mr

15.36 & 7.68 & 4.72$\times 10^{-14}$ & 0.1268 & 0.1141		& 3.24$\times 10^{-13}$\\

15.36 & 3.84 & 1.35$\times 10^{-14}$ & 0.1268 & 0.1232  	& 1.15$\times 10^{-12}$\\

19.20 & 7.68 & 4.72$\times 10^{-14}$ & 0.1025 & 0.0954		& 5.84$\times 10^{-13}$\\

\br
\end{tabular}
\end{table}

Qubits are initialized in the $\left|0\right\rangle$ state by applying an electrostatic potential~$\Delta V_\imath(t)$ so that the left~DB is lower in energy thus attracting the pair's excess electron~\cite{barrett1}. 
When initialization is complete, the electrostatic bias is eliminated, and tunneling between the two DBs commences. A lattice deformation due to charge localization as in Fig.~\ref{fig:qubit}(b) is present during subsequent tunneling, but is expected to relax at a much lower rate than $\Delta$ (by a few orders of magnitude), hence having a small decoherence effect.
Application of a static potential has shown polarization to be achievable~\cite{HBD+09}. In the same experiment, steps towards qubit-specific readout were achieved by STM detection of the excess charge preferentially localized at one site in a DB-DB$^-$ pair.
This experiment thus shows that both state preparation on one side and readout of $\left|0\right\rangle$ vs $\left|1\right\rangle$ state is feasible. 
%
Fast readout would be desirable, not only for error correction, but also to measure decoherence. One approach to fast readout is to couple the charge qubit to single electron transistor (SET) and detect the changes in its output when the qubit is in the $\left|0\right\rangle$ as opposed to the $\left|1\right\rangle$ state. The charging state of a DB was shown to affect the STM current through  a nearby molecule attached to the surface~\cite{PLP+05}: the molecule's electronic  structure can be Stark-shifted by the DB's excess electron.

Hamiltonian~$H_{\rm{q}}$ enables a universal set of gates~\cite{BBC+95}. Single- and two-qubit gates are effected by varying the inter-dot tunneling rate by tilting the potential landscape then rapidly turning off the tilting. Such fast and spatially precise control is beyond the current capability of standard electronics, but is in principle conceivable by placing a suitable pattern of metallic nanowires near the surface and irradiating it with a laser pulse. The resulting electromagnetic field, created via plasmonic action~\cite{MA05},  can bias the surface with a temporal control comparable to the duration of the pulse, which can be as short as femtoseconds. 
The laser carrier frequency should be low enough to avoid charging and discharging of DBs through excitation processes, thereby causing qubit losses.
Different gates could be effected by time-varying biases achieved by controlled laser pulses. 

Scalability of our surface charge-qubit quantum computer follows the same arguments as for those cases, but of course better understanding of small-scale devices is required to assess scalability to many-qubit devices. An important feature in our DB system is that qubit cross-talk is minimized by screening effects in the semiconductor.
At this early stage, bearing in mind that many implementation details are in need of development, possible computing schemes appear to  be: a four-rail flying qubit model analogous to the one for electron-spin qubits in bulk silicon~\cite{HGFW06}, or a one-way quantum computer~\cite{RB01}, where the qubits are stationary.

\section{Summary and Outlook}
We show that closely-separated DB-DB$^-$ pairs on the silicon surface behave as charge qubits, with excellent coherence properties following the extreme miniaturization of qubits, indeed to the atomic realm.
This is a consequence of the fact that the tunneling rate is extremely high due to atomic-scale proximity of DBs, whereas the major source of decoherence scales weakly with separation. The scaling advantage comes at the price of having to achieve rapid gating control. As far as we can see, such a scheme entails some technical objectives to be achieved: scaling down the nanowire network needed for biasing qubits; accurate control of the amplitudes of the pulsed fields; ways to incorporate readout during the computation for the purpose of quantum error correction.

A logical step forward would be to further develop quantum computing implementations using our DB-DB$^-$ qubits, with the need of addressing all DiVincenzo's criteria for such architectures, which will require much elaboration.
Near-future efforts will be concentrated on developing ways to investigate experimentally a small numbers of DBs. In the first instance, experimental characterization of the decoherence for these charge qubits is of paramount importance. Whereas time-domain control is ultimately required, decoherence can be studied in the near term by fluorescence techniques: charge qubits are dipoles that will fluoresce in the THz regime, and decoherence can be extracted from linewidths. 


In addition to weak decoherence, our scheme has another important advantage over other semiconductor charge qubit proposals: the charge qubits are on the surface rather than in the bulk medium, thus enabling more direct preparation, control, and readout. Some of the required DB quantum dot dynamics have already been demonstrated.  We believe the findings outlined here could reinvigorate charge qubit prospects for quantum computing.

\ack{
This project has been supported by NSERC, MITACS, QuantumWorks, \emph{i}CORE.
RAW and BCS are CIFAR Fellows.
PX acknowledges support from the National Natural Science Foundation of China, Grant No. 10944005
and the Southeast University Startup Fund.}

\section*{References}

\bibliography{qubit_NJP_refs}{}
\bibliographystyle{unsrt}

\end{document}